\documentclass[aps,superscriptaddress,preprint]{revtex4}
\usepackage[dvips]{graphicx}
\usepackage{latexsym}
\usepackage{natbib}

\begin{document}
\title{Fractal Boundaries of Complex Networks}

\author{Jia Shao$^{1}$, Sergey V. Buldyrev$^{2,1}$, Reuven Cohen$^{3}$, 
Maksim Kitsak$^{1}$, Shlomo Havlin$^{4}$, and H. Eugene Stanley$^{1}$}

\affiliation{$^1$Center for Polymer Studies and Department of Physics,
  Boston University, Boston, Massachusetts 02215, USA\\
$^2$Department of Physics, Yeshiva University, 500 West 185th Street, New York, New York 10033, USA\\
$^3$Department of Mathematics, Bar-Ilan 
University, 52900 Ramat-Gan, Israel \\
$^4$Minerva Center and Department of Physics, Bar-Ilan University, 52900
Ramat-Gan, Israel \\}

\date{ \today}

\begin{abstract} 

We introduce the concept of boundaries of a complex network as the set of nodes 
at distance larger than the mean distance from a given node in the network.
We study the statistical properties of the boundaries nodes of complex
networks.
We find that for 
both Erd\"{o}s-R\'{e}nyi and scale-free model networks, as well as for several real networks,
the boundaries
have fractal properties. In particular, 
the number of boundaries nodes {\it B} 
follows a power-law probability density
function which scales as $B^{-2}$. 
The clusters formed by the boundary nodes are 
fractals with a fractal dimension $d_{f} \approx 2$. 
We present analytical and numerical evidence supporting these results 
for a broad class of networks. Our findings imply potential applications for epidemic spreading.

\end{abstract}

\maketitle
Many complex networks are ``small world'' due to the very small 
average distance {\it d} between two randomly chosen nodes. 
Usually $d \sim \ln N$, 
where $N$ is the number of nodes \cite{er1,bollo,bararev,milgram,ws,cohena}. 
Thus, starting from a randomly chosen node following the shortest 
path, one can reach any other node in a very small number of steps.
This phenomenon is called ``six degrees of separation'' in social networks \cite{milgram}. 
That is, for most pairs of randomly chosen people, the shortest ``distance'' between them is not more than six.  
Many random network models, such as Erd\"{o}s-R\'{e}nyi 
network (ER) \cite{er1}, Watts-Strogatz network (WS) \cite{ws} and random 
scale-free network (SF) \cite{bararev,cohena,mendes,vespig}, as well as many real networks,
have been shown to possess this small-world property. 

Much attention has been devoted to the structural properties of networks within the average
distance $d$ from a given node. However, almost no attention
has been given to nodes which are at distances greater than $d$ from a given node. 
We define these nodes as the boundaries of 
the network and study the ensemble of boundaries. 
An interesting question is how many ``friends of friends of friends
etc...'' one has at distance greater than the average distance $d$?  
What is their probability distribution and what is the structure 
of the boundaries?
 The boundaries have an important role in several scenarios, such as in the  
spread of viruses or information in a human social network. 
If the virus (information) spreads from one node to all its
nearest neighbors, and from them to all next nearest neighbors and
further on until $d$, 
how many nodes do not get the virus (information), and what is 
their distribution with respect to the origin of the infection. Our
results may explain why epidemics such as ``black death'' in medieval Europe stopped before
reaching the entire population.


In this Letter, we find theoretically and numerically that the nodes at the boundaries, which are of 
order $N$, exhibit similar fractal features for many types of  
networks, including ER 
and SF models as well as several real networks. Song {\it et al.} \cite{song1} found that some networks have
fractal properties while others do not. Here we show that almost all
model and real networks 
including non-fractal networks have fractal
features at their boundaries which are different from Song {\it et al}.

\begin{figure}[h!]
  \centering
  \includegraphics[width=7.5cm,angle=0]{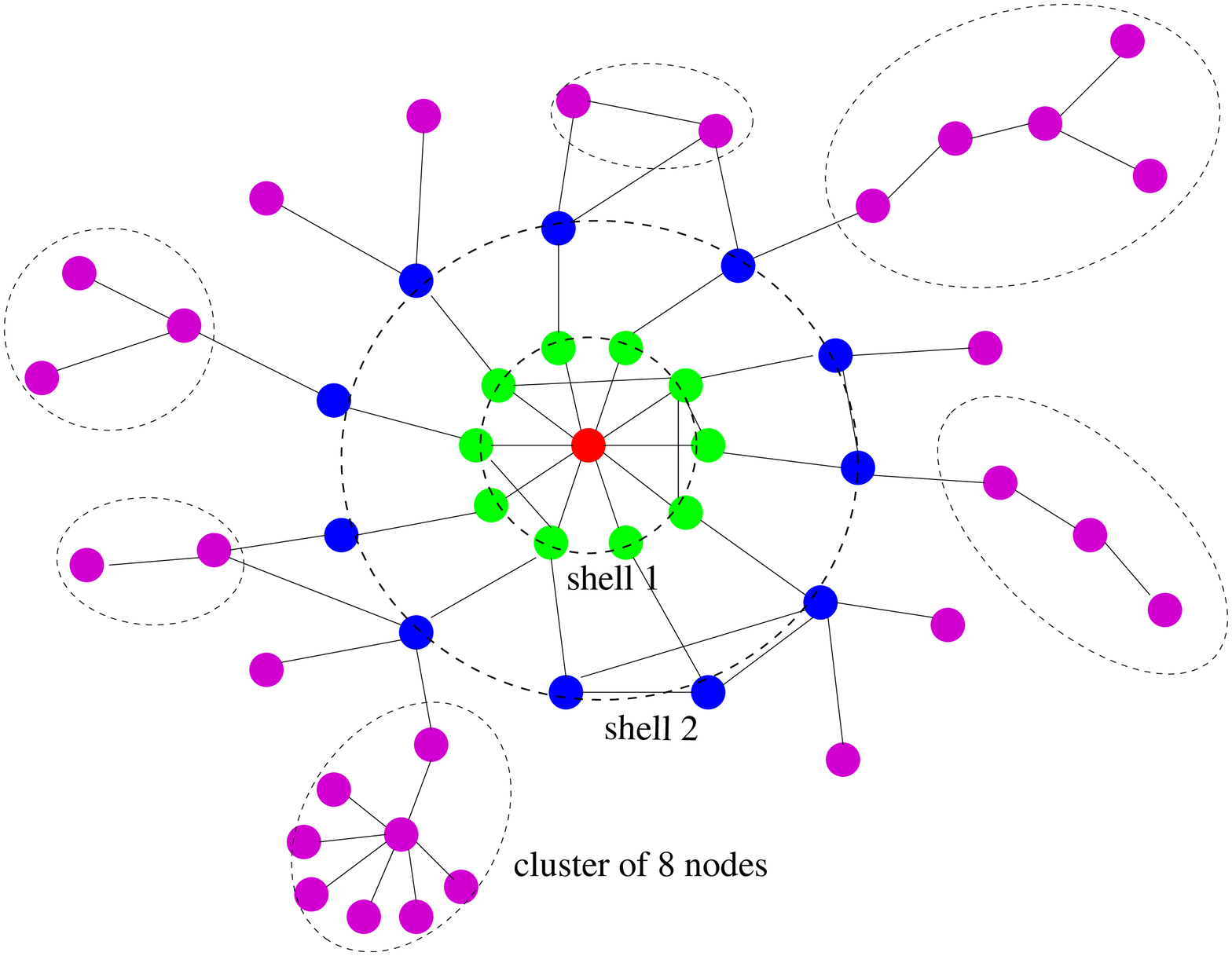} 
\caption{ (Color on line) Illustration of shells and clusters originating from a randomly chosen node, 
which is shown in the center (red).
Its neighboring nodes 
are defined as shell 1, the nodes at distance $\ell$ are defined as shell $\ell$.
When removing all nodes with $\ell <3$, the remaining network becomes fragmented into 12 clusters. 
   }
\label{fig1}
\end{figure}

Fig.~\ref{fig1} demonstrates our approach and analysis. 
For each node, we identify the nodes at 
distance $\ell$ from it as nodes in shell $\ell$. 
We chose a random origin node
 and count the number of nodes {\it $B_{\ell }$} at shell $\ell$. 
We see that {\it $B_{1}$}=10, {\it $B_{2}$}=11,
{\it $B_{3}$}=13, etc... We estimate the average distance $d\approx 2.9$ 
by averaging the distances between all pairs of nodes. 
After removing nodes with $\ell <d=2.9$, the network is fragmented into 12 clusters,  
with sizes $s_{3}$=$\{1,1,2,5,1,3,1,1,8,1,2,3\}$.

\begin{figure*}[htp]
    \includegraphics[width=6cm,angle=-90]{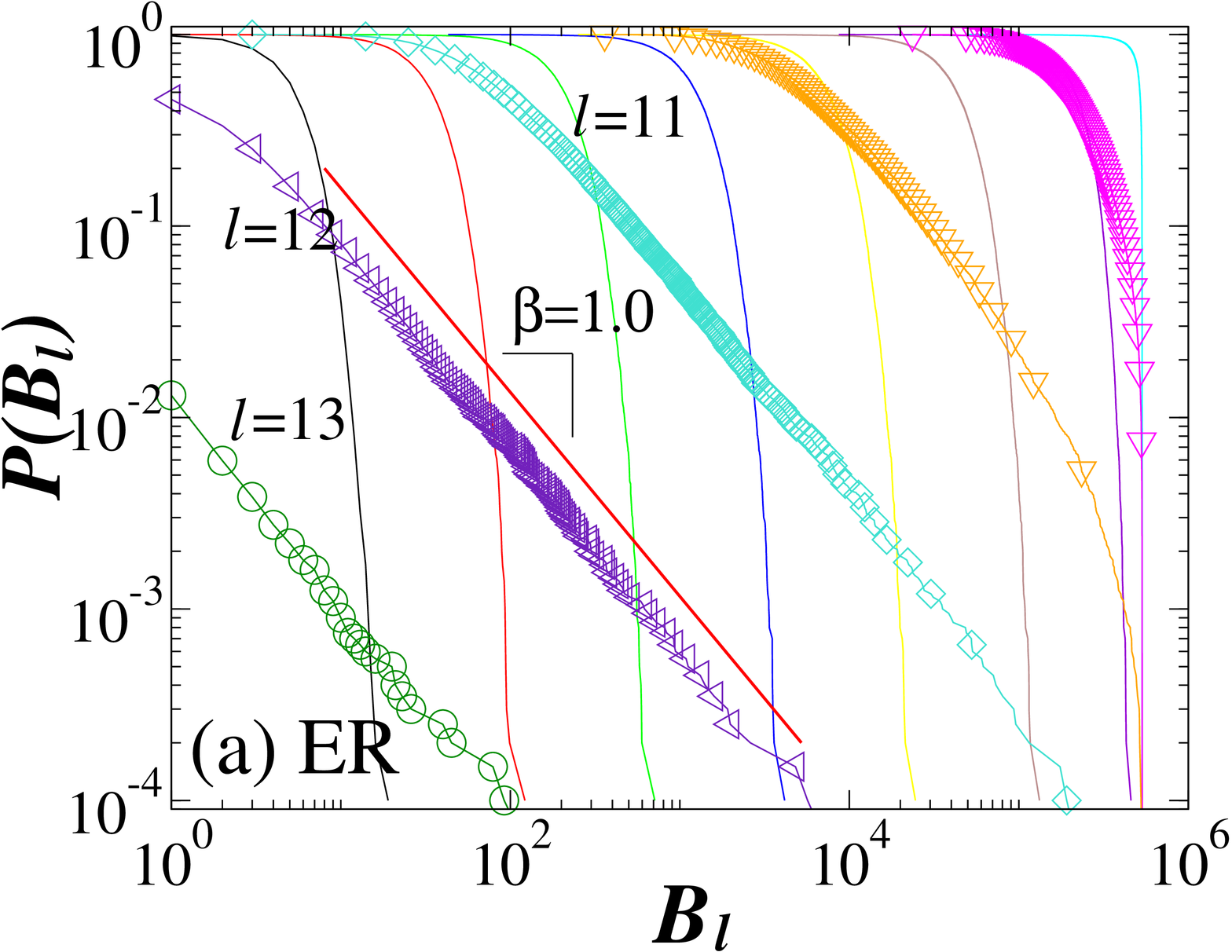}
   \includegraphics[width=6cm,angle=-90]{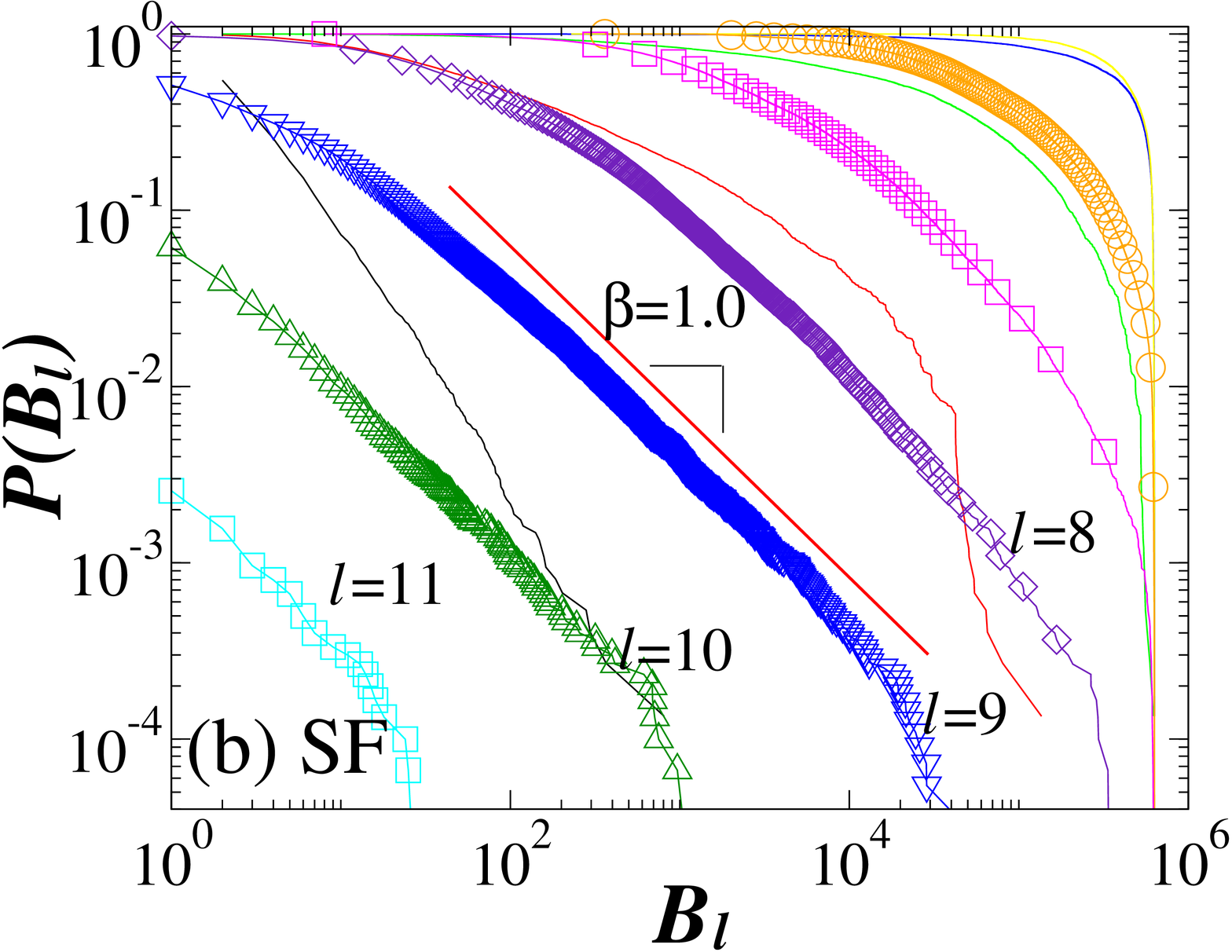}
   \includegraphics[width=6cm,angle=-90]{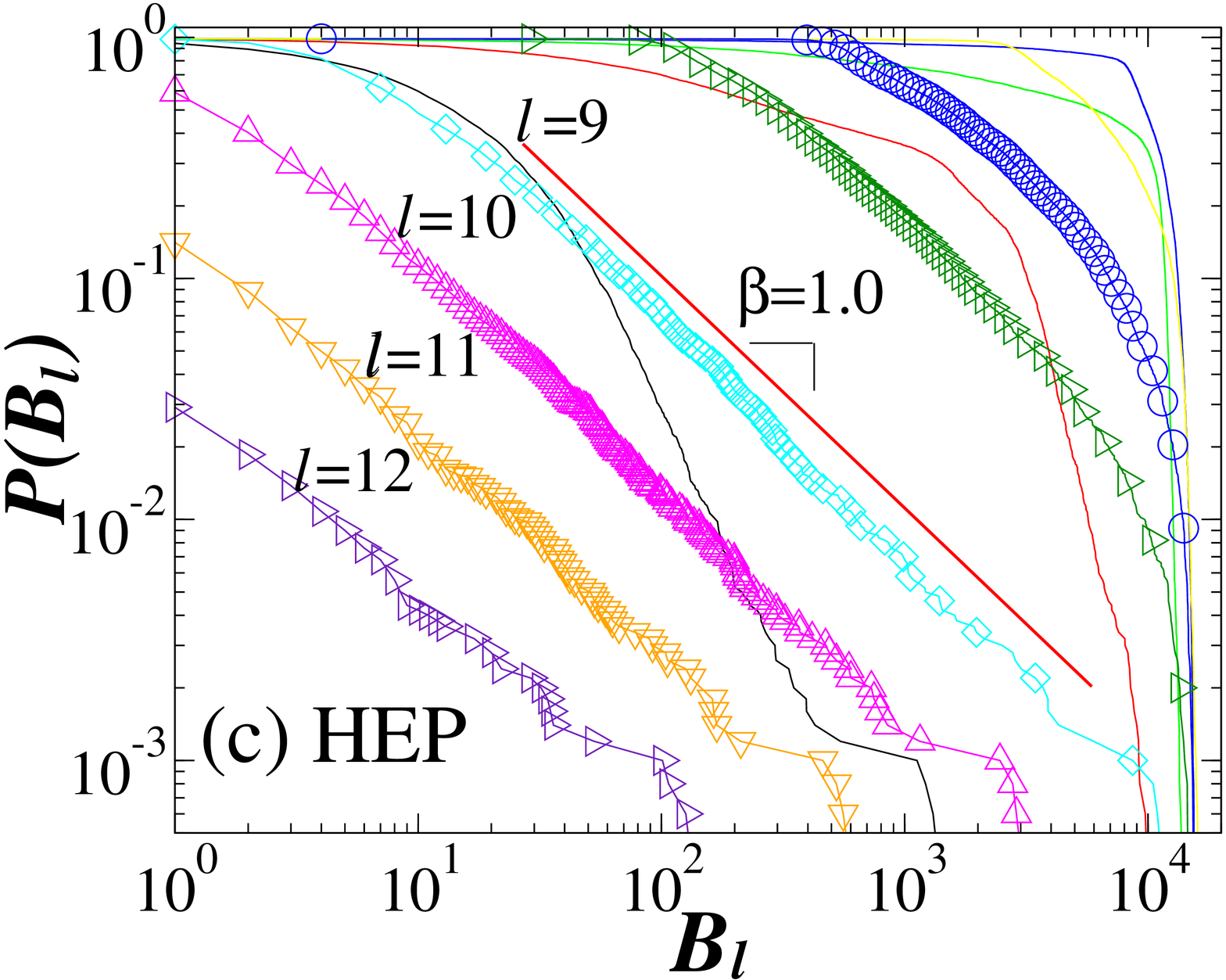}
   \includegraphics[width=6cm,angle=-90]{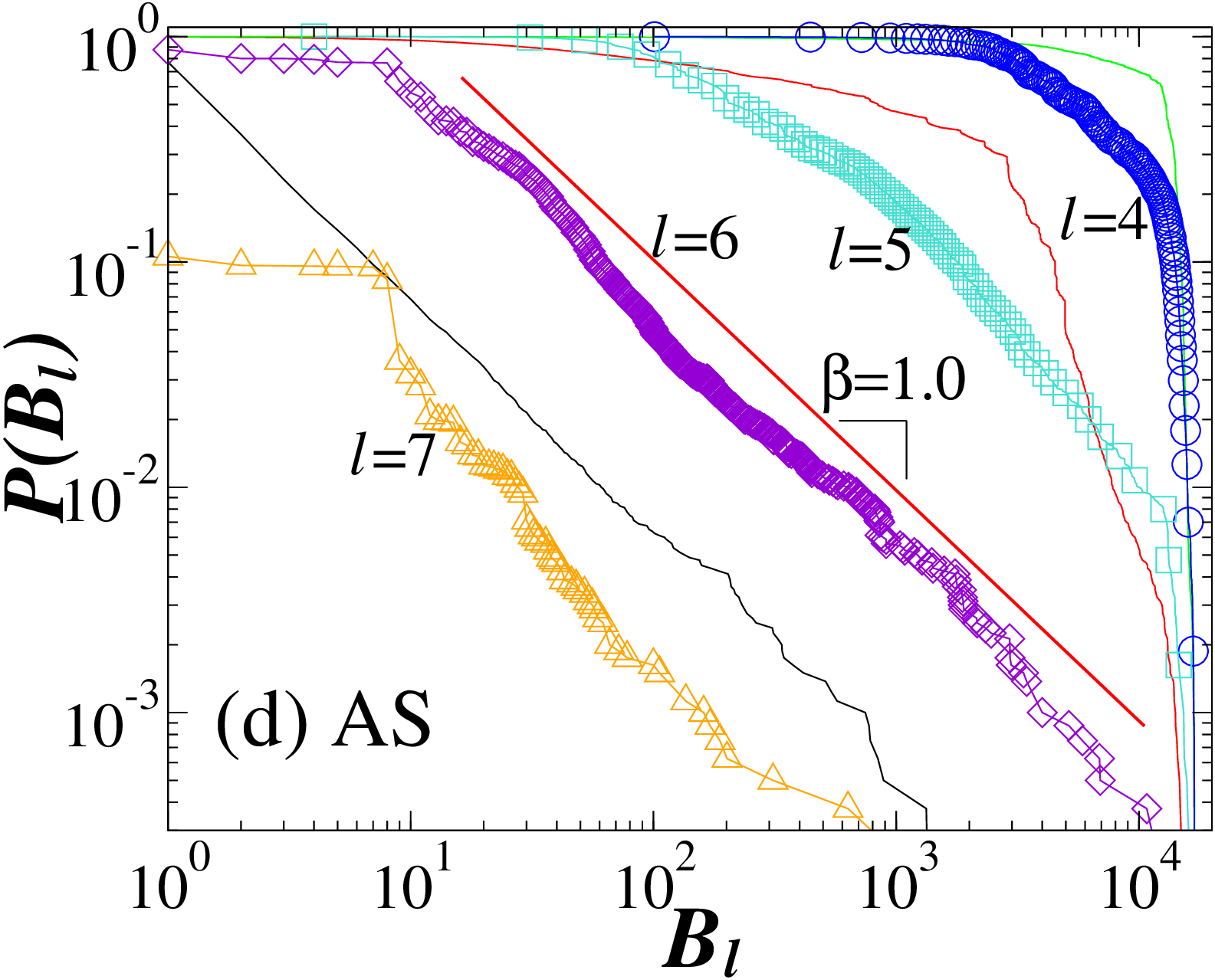}
 \caption{The cumulative distribution function, $P(B_{\ell})$, for two random network models: 
(a) ER network with $N=10^{6}$ nodes and $\langle k\rangle=6$, 
and (b) SF network with $N=10^{6}$ nodes and $\lambda=2.5$, and two real networks:
(c) the High Energy Particle (HEP) physics citations network and (d) the Autonomous System (AS) Internet network.
The shells with $\ell>d$ are marked with their shell number.
The thin lines from left to right represent shells 
$\ell=$1, 2... respectively, with $\ell<d$. For $\ell>d$, 
$P(B_{\ell})$ follows a power-law distribution 
$P(B_{\ell}) \sim B_{\ell}^{-\beta}$, 
with $\beta\approx 1$  
(corresponding to $\tilde{P}(B_{\ell})\sim B_{\ell}^{-2}$ for the pdf).
The appearance of a power law decay only happens for $\ell$
larger than $d \approx 7.9$ for ER and $d \approx 4.7$ 
for the SF network. The straight lines represent a slope of $-1$.
}
\label{fig2}
\end{figure*}

We begin our study by simulating ER 
and SF networks, and later present 
analytical proofs. Fig.~\ref{fig2}a shows simulation results for the
number of nodes $B_{\ell}$ reached from a 
randomly chosen origin node for an ER 
network. The results shown are for a single network realization of size $N=10^{6}$, with
average degree $\langle k \rangle=6 $ and $d\approx 7.9$ \cite{similar}.
For $\ell <d$, the cumulative distribution function, $P(B_{l})$, which is the probability
that shell $\ell$ has more than $B_{\ell}$ nodes, decays exponentially for $B_{\ell}>B_{\ell}^{\ast}$, 
where $B_{\ell}^{\ast}$ is the maximum typical size of shell $\ell$ \cite{fig3c}.
However, for $\ell >d$, we observe a clear transition to a power law decay behavior, where 
$P(B_{\ell})\sim B_{\ell}^{-\beta}$, with $\beta\approx 1$ and the 
pdf of $B_{\ell}$ is $\tilde{P}(B_{\ell})\equiv d P(B_{\ell})/d B_{\ell}\sim B_{\ell}^{-2}$.
Thus, our results suggest a broad ``scale-free'' distribution for the number of nodes 
at distances larger than $d$. 
This power law behavior demonstrates the fractal nature of the boundaries of network, suggesting
that there is no characteristic size and a broad range of sizes can
appear in a shell at the boundaries. Further
fractal features of the boundaries structure will be shown below.

In SF networks, 
the degrees of the nodes, {\it k}, follow a power law distribution function
$q(k) \sim k^{- \lambda}$, 
where the minimum degree of the network is chosen to be 2.
Fig.~\ref{fig2}b shows, for SF networks with $\lambda=2.5$, similar power law 
results, $P(B_{\ell})\sim B_{\ell}^{-\beta}$ for $\ell>d$ as for ER, 
with a similar power $\beta\approx 1$.
We find similar results also for $\lambda > 3$ (not shown).

To test how general is our finding, we also study several 
real networks (Figs.~\ref{fig2}c, 2d), including the 
High Energy Particle (HEP) physics citations network \cite{hep} and 
the Autonomous System (AS) Internet network \cite{dimes,shai}. 
Our results suggest that 
the fractal properties of the boundaries appear also in both networks, with similar values of $\beta\approx 1$ 
for $\ell>d$ \cite{yeast}.

\begin{figure*}[htp]
  \includegraphics[width=6cm,angle=-90]{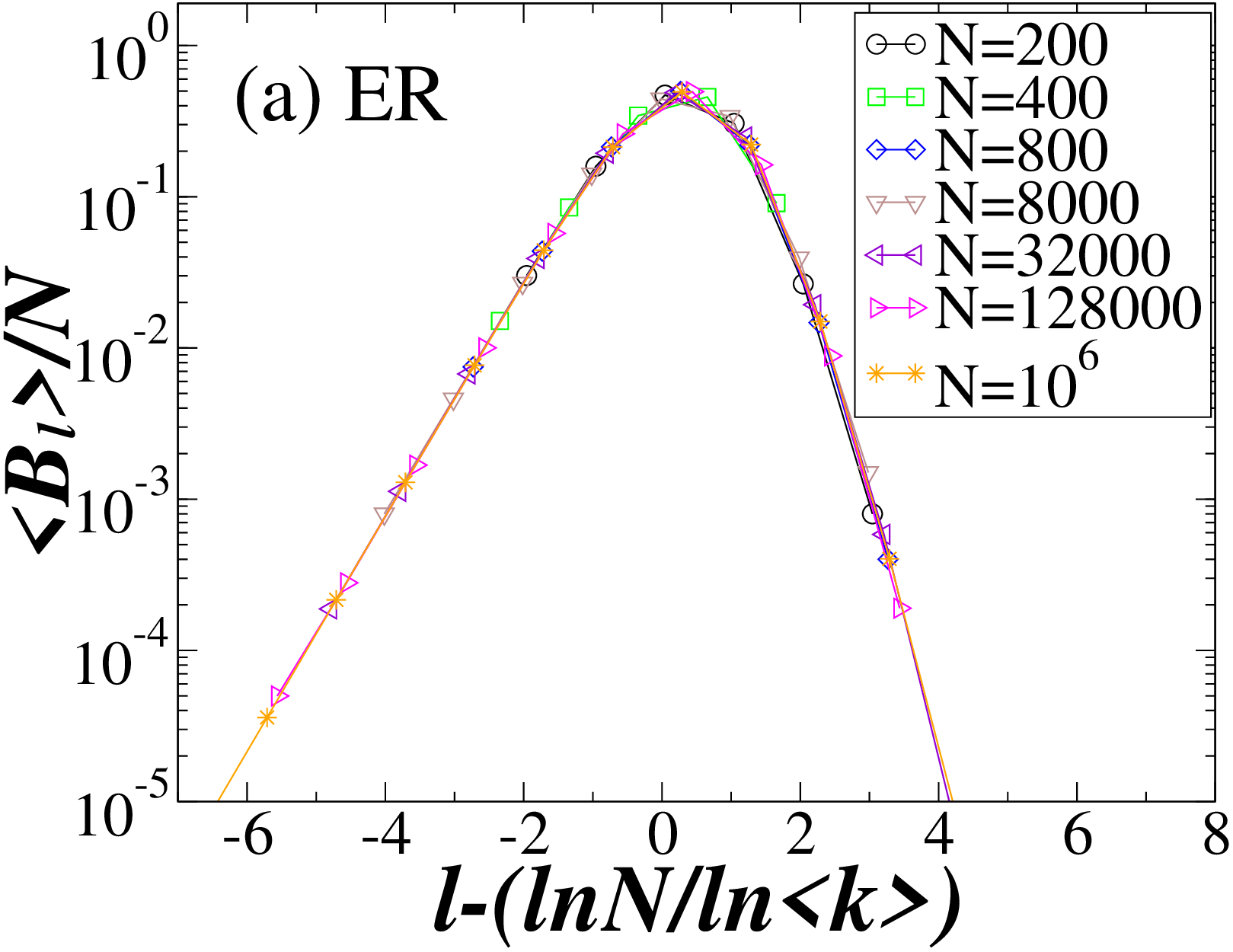}
  \includegraphics[width=6cm,angle=-90]{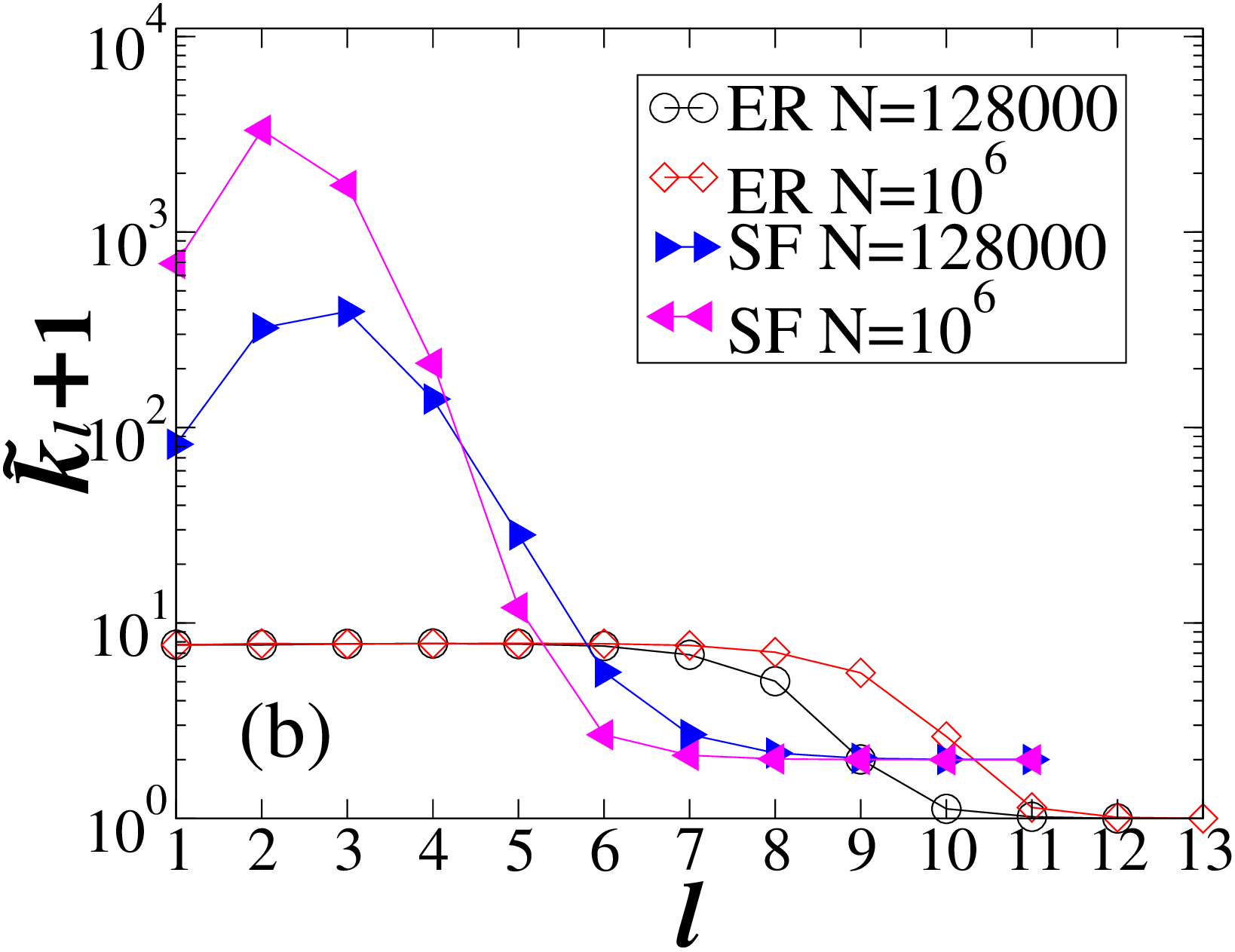}
   \includegraphics[width=6cm,angle=-90]{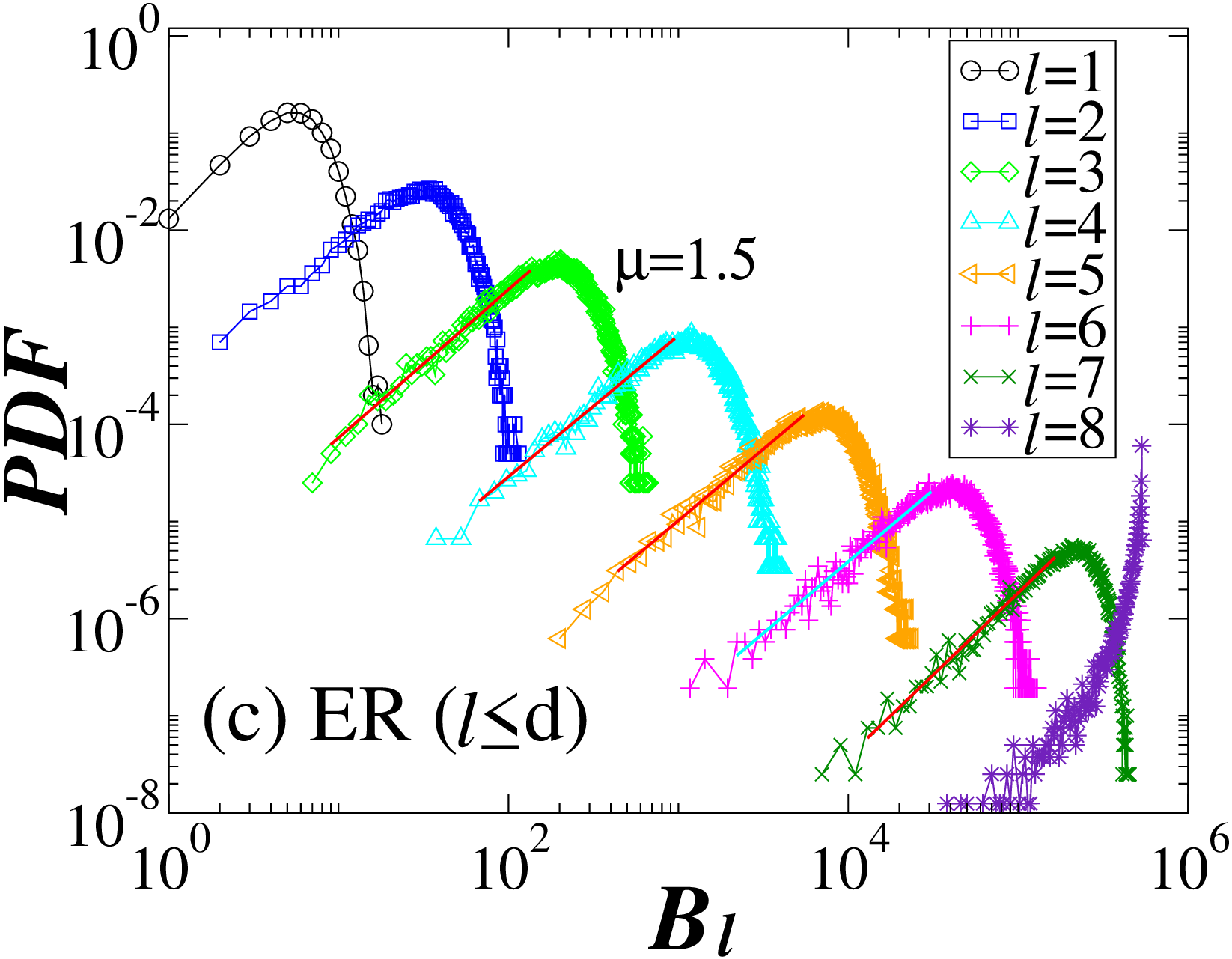}
  \includegraphics[width=6cm,angle=-90]{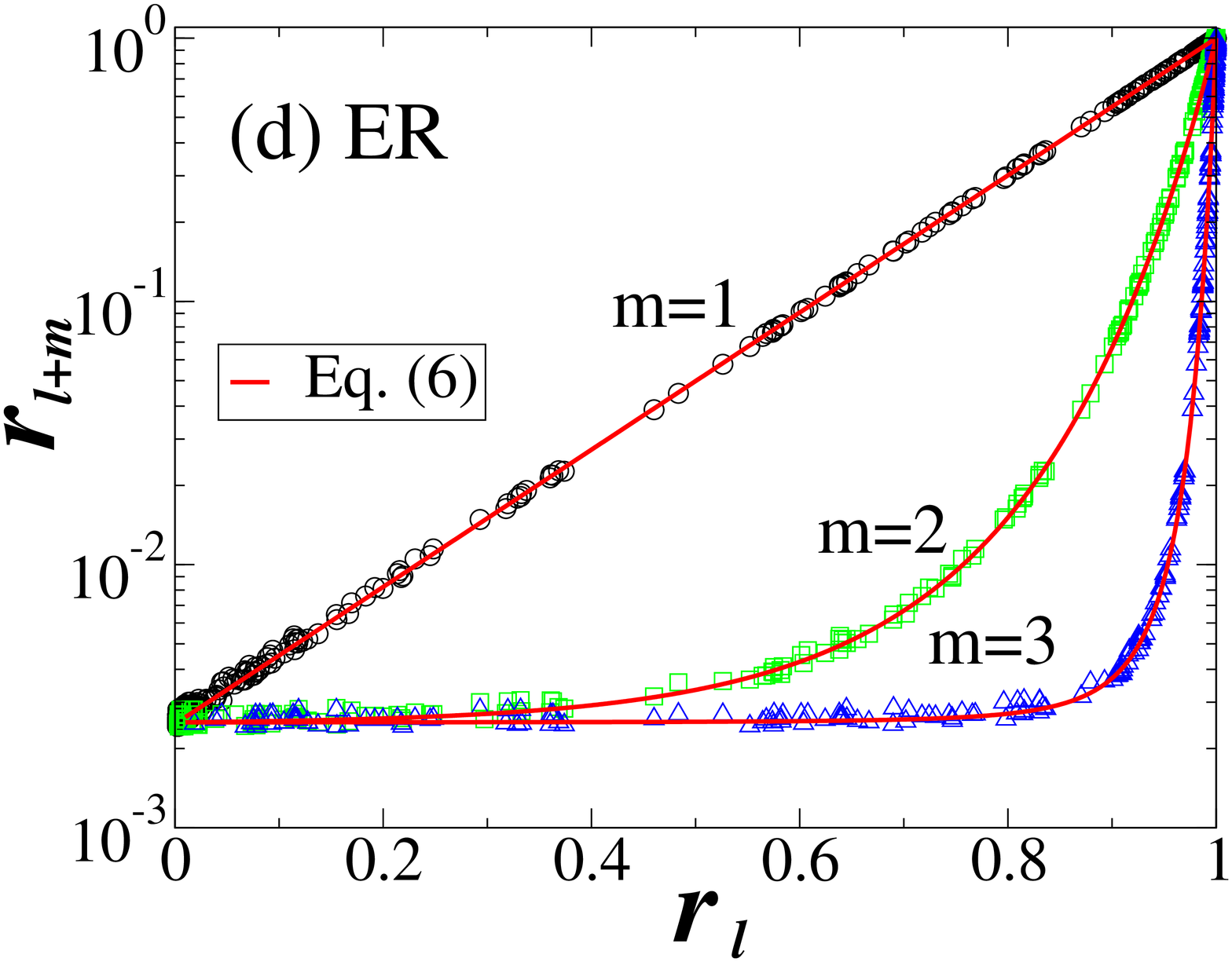}
    \caption{(Color on line) (a) Normalized average number of 
nodes at shell $\ell$, $\langle B_{\ell} \rangle /N$,  
as a function of $\ell-\ln N/\ln \langle k \rangle$ for 
ER network with $<k>$=6. For different $N$, the curves collapse.
(b) $\tilde{k}_{\ell}+1$, which is $\langle k_{\ell}^{2}\rangle / \langle k_{\ell} \rangle$, 
    as function of $\ell$ shown 
    for both ER and SF network.
(c) The probability distribution function $\tilde{P}(B_{\ell})$ 
    in shells $\ell\leq d$ for ER network. 
    For small values of $B_{\ell}$, 
   $\tilde{P}(B_{\ell})\sim B_{\ell}^{\mu}$, where $\mu$ 
    depends on the $\langle k \rangle$ of the network (Eq. (4)). 
(d) The fraction of nodes outside shell $\ell +m$, $r_{\ell +m}$, 
as a function of $r_{\ell}$ for ER network, 
where $r_{\ell}$ is calculated for any possible $\ell$.
The (red) lines represent the theoretical iteration function (Eq. (6)).  
}
\label{fig3}
\end{figure*}

Next we ask how many nodes are on average at the boundaries? Are they a finite fraction of $N$, or less? 
In Fig.~\ref{fig3}a, we study the mean number 
$\langle B_{\ell}\rangle$ in shell $\ell$, and plot $\langle B_{\ell}\rangle /N$ as function of 
$\ell-\ln N /\ln \langle k \rangle$ for different values of $N$ for ER network.
The term $\ln N /\ln \langle k \rangle$ represents the average distance {\it d} of the network \cite{bollo}.
We find that, for different values of $N$, the curves collapse, supporting
a relation independent of network size $N$. Since $\langle B_{\ell}\rangle / N$ is 
apparently constant and 
independent of $N$, it follows that $\langle B_{\ell}\rangle \sim N$,
i.e., a finite fraction of {\it N} nodes appear at each shell including shells with $\ell>d$. 
We find similar behavior for SF network with $\lambda=3.5$ (not shown).
The branching factor \cite{cohen} of the network is $\tilde{k}=\langle k^{2}\rangle / \langle k \rangle-1$,
where the averages are calculated for the entire network.
Similarly, we define $\tilde{k}_{\ell}=\langle k_{\ell}^{2}\rangle / \langle k_{\ell} \rangle-1$, where 
the averages are calculated only for nodes in shell $\ell$. 
Above the average distance, $\tilde{k}_{\ell}+1$
decreases with $\ell$ for both ER and SF networks (Fig.~\ref{fig3}b). 
Thus, at the shells where power law behavior of $P(B_{\ell})$ appears (Fig.~\ref{fig2}), 
the nodes have much lower $\tilde{k}_{\ell}+1$
compared with the entire network. The approach of $\tilde{k}_{\ell}+1$ to 
1 (ER network) and 2 (SF network) 
is consistent with a critical behavior at the boundaries of the network \cite{cohen}.

Fig.~\ref{fig3}c shows that $\tilde{P}(B_{\ell})$ for $\ell<d$ and small values 
of $B_{\ell}$ increase as a power law,
$\tilde{P}(B_{\ell})\sim B_{\ell}^{\mu}$, for ER network, where $\mu$ depends on $\tilde{k}$ 
(supporting the theory developed below). We
define the fraction of nodes outside shell $m$ as 
$r_m=1-(\sum_{\ell=1}^m B_\ell)/N$.
There exists a functional relation Eq. (6), which is independent of $\ell$, between 
any two $r_{\ell}$ and $r_{\ell+m}$ ($m=1, 2, 3...$), 
for ER network in Fig.~\ref{fig3}d.  
Figs.~\ref{fig3}c, 3d provide empirical evidences for the theory developed below.

\begin{figure*}[htp]
     \includegraphics[width=6cm,angle=-90]{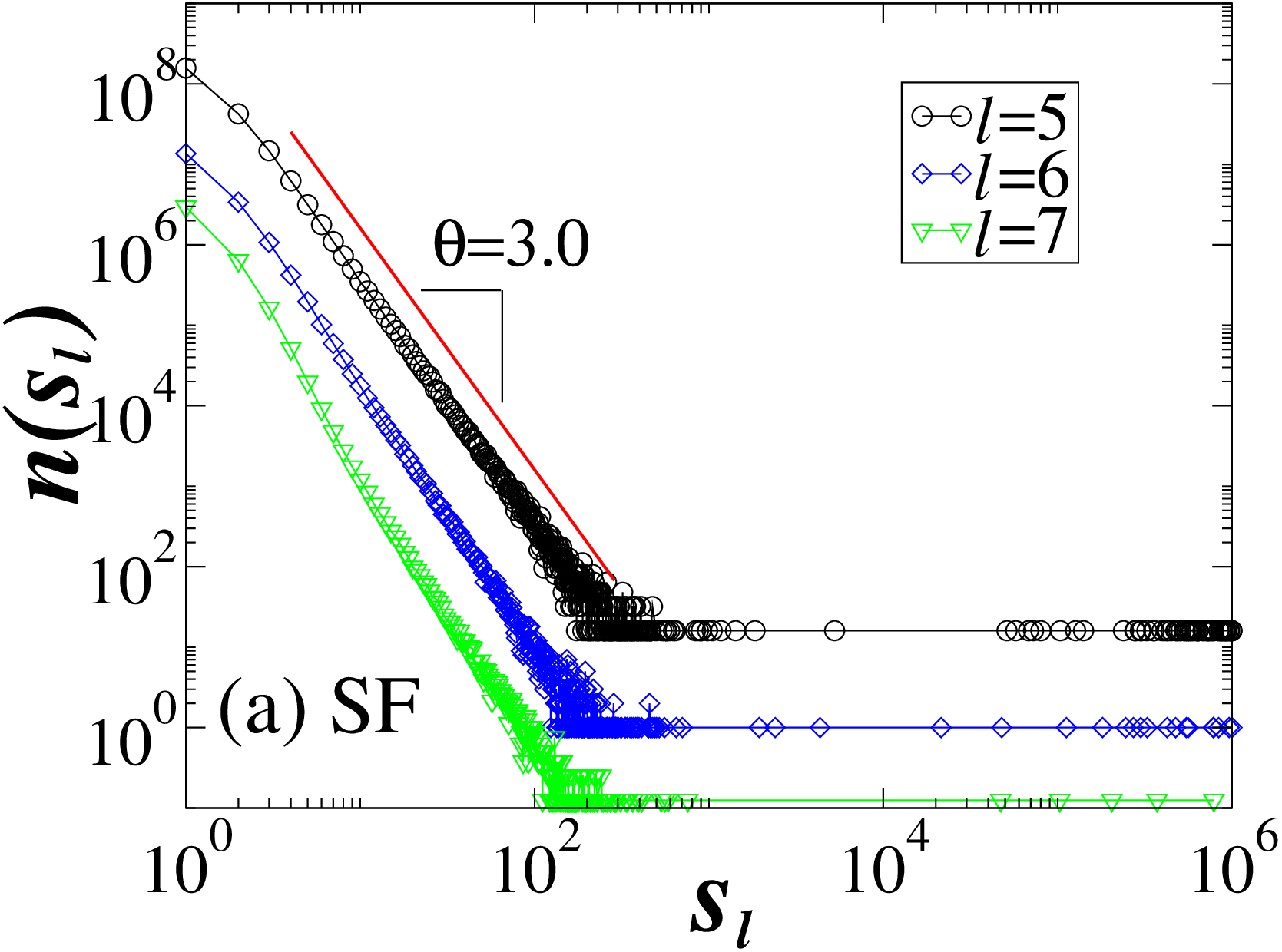}
       \includegraphics[width=6cm,angle=-90]{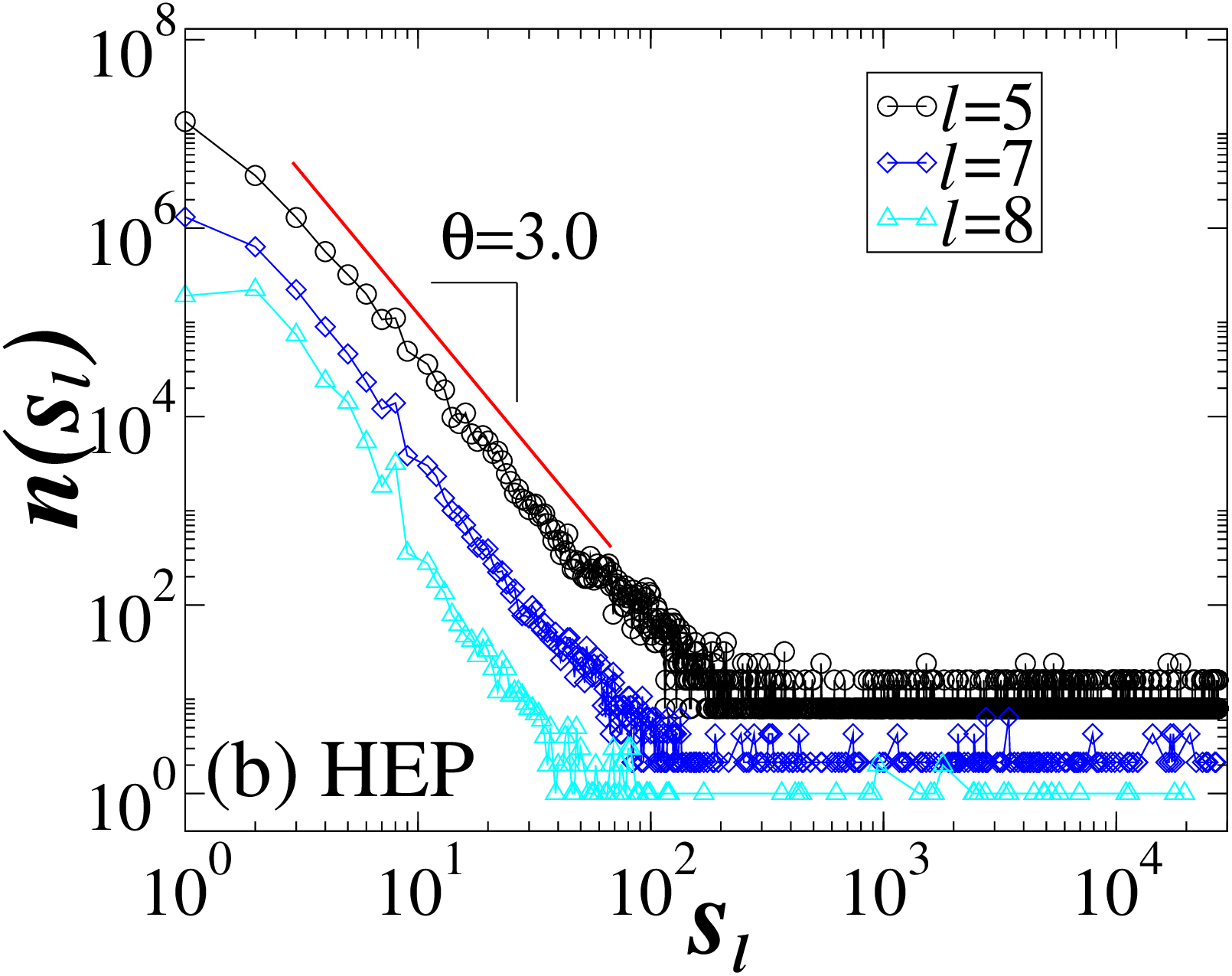}  
      \includegraphics[width=6cm,angle=-90]{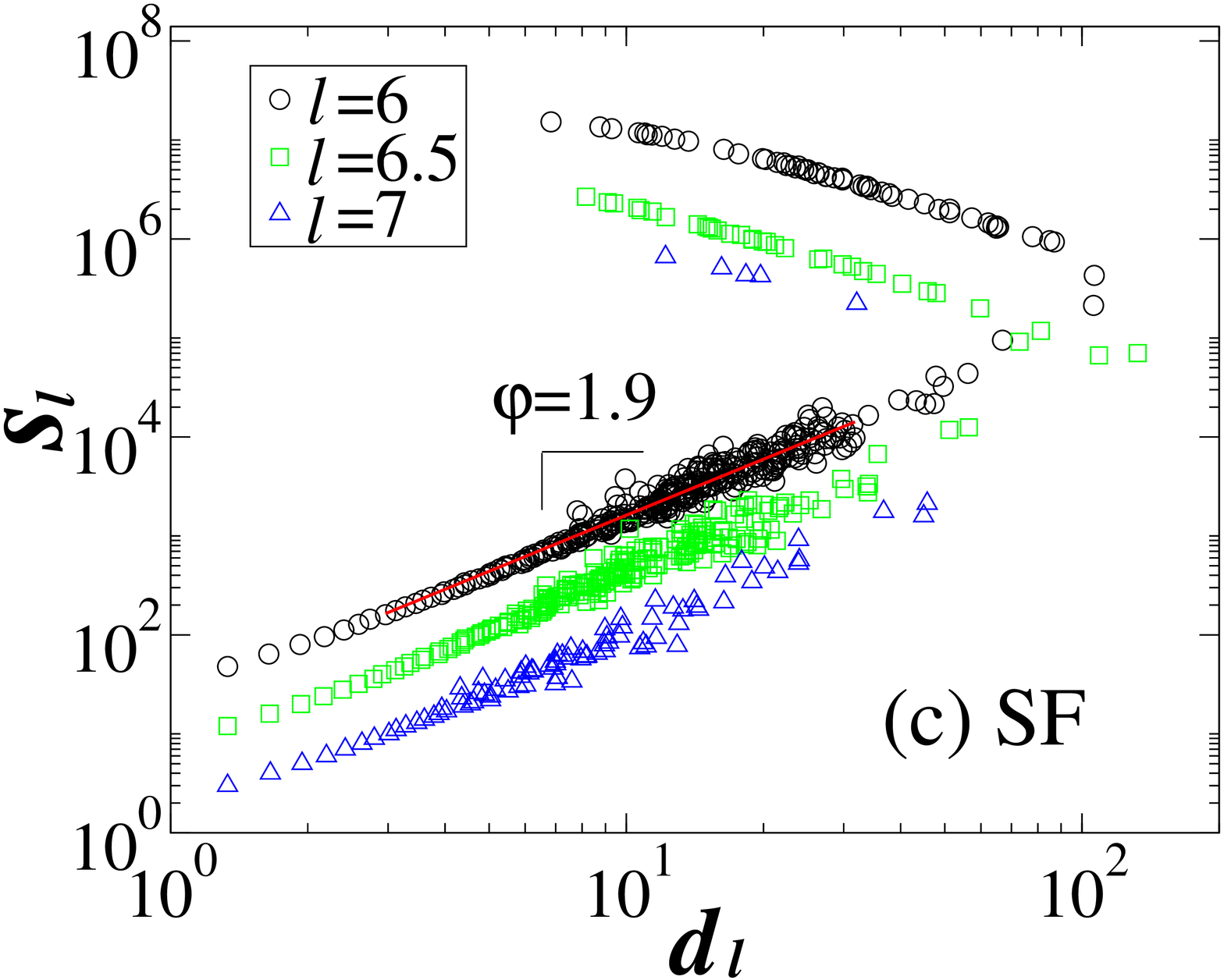}
       \includegraphics[width=6cm,angle=-90]{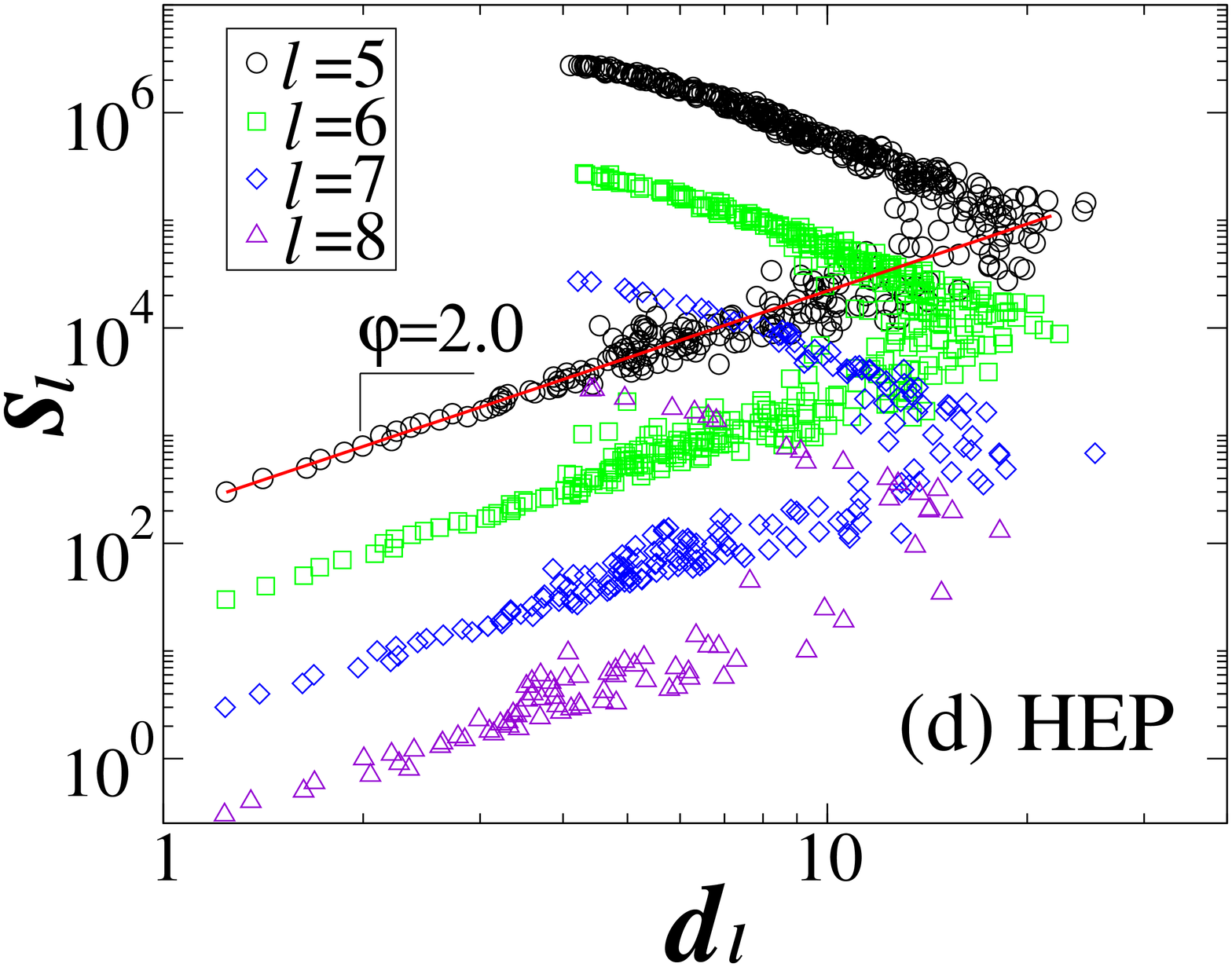}
 \caption{ The number of clusters of sizes $s_{\ell}$, $n(s_{\ell})$, as function of $s_{\ell}$
after removing nodes within shell $\ell$ for:  
(a) SF network with $N=10^{6}$ and $\lambda=2.5$, (b) HEP citations network, and $s_{\ell}$ as function of 
average distance $d_{\ell}$ of the clusters for (c) SF network with N=$10^{6}$ and $\lambda=2.5$, 
(d) HEP citations network. 
The relation between $n(s_{\ell})$ and $s_{\ell}$ is characterized by a power law, $n(s_{\ell})\sim s_{\ell}^{-\theta}$, with $\theta \approx 3$. 
Also, $s_{\ell}$ scales with $d_{\ell}$ as $s_{\ell} \sim d_{\ell}^{\varphi}$, 
with $\varphi\approx 2 $.
}
\label{fig4}
\end{figure*}

Next, we study the structural properties of the boundaries.
Removing all nodes that are within a distance $\ell>d$ (not
including shell $\ell $), the network will become fragmented into several clusters (see Fig.~\ref{fig1}). 
We denote the size of those clusters as $s_{\ell}$, the number of clusters 
of size $s_{\ell}$ as $n(s_{l})$, 
and the average distance 
in the clusters as $d_{\ell}$ \cite{fraction}. 
We find $n(s)\sim s^{-\theta}$, with $\theta \approx 3.0$ (Figs.~\ref{fig4}a and 4b). Similar 
relations are also found for ER and other real networks. 
The relation between the size of the clusters $s_{\ell}$ and their mean distance
$d_{\ell}$ is shown in Figs.~\ref{fig4}c and 4d, for SF ($\lambda=2.5$) 
and HEP citations networks respectively. These plots suggest 
a power law relation, $s_{\ell} \sim d_{\ell}^{\varphi}$, 
with $\varphi\approx 2$. It indicates that the clusters
at the boundaries are fractals with fractal dimension $d_f=2$ as
percolation clusters at criticality \cite{cohenpre}. 
Note that, for very large 
clusters their average distances $d_{\ell}$ decrease with size, 
suggesting that the largest clusters are not fractals. We find that the fractal 
dimension is $d_{f}=\varphi \approx 2$ also for ER, 
SF with $\lambda=3.5$ and several other real networks.

Next we present analytical derivations supporting the above numerical results. 
We denote the degree distribution of a network as $q(k)$. 
In infinitely large network we can neglect loops for $\ell<d$ and approximate 
the behavior of forming a network as a 
branching process \cite{Harris1,bingham,braunstein}. 
The probability of reaching 
a node with $k$ outgoing links 
through a link is $\tilde q(k)=(k+1)q(k+1)/\langle k\rangle$.
 The probability of number of branches equals to $\tilde{q}(k)$.
We define the generating function of $q(k)$ as 
$G_{0}(x)\equiv \sum _{k=0}^{\infty} q(k)x^k$, the
generating function of $\tilde{q}(k)$ as
$G_{1}(x)=\sum _{k=0}^{\infty}\tilde q(k)x^k=G_{0}^{'}(x)/\langle k\rangle$. 
For ER networks we have $G_{0}(x)=G_{1}(x)=e^{\langle k \rangle(x-1)}$. 
The generating function for the number of nodes, $B_{m}$,
at the shell $m$ is \cite{mnewman}: 
\begin{equation}
\tilde{G}_{m}(x)=G_{0}(G_{1}(...(G_{1}(x))))=G_{0}(G_{1}^{m-1}(x)),
\end{equation}
where $G_{1}(G_{1}(...)) \equiv G_{1}^{m-1}(x)$ is the result of 
applying $G_{1}(x)$, $m-1$ times.
$\tilde P(B_m)$, which is the probability distribution of $B_m$, is the coefficient of $x^{B_{m}}$ in the Taylor
expansion of $\tilde{G}_m(x)$.

For shells with large $m$ but still much smaller than $d$, we expect \cite{mnewman}
that the number of nodes will increase by a factor of $\tilde k$. 
It is possible to show \cite{bingham} that $G_{1}^{m-1}(x)$ 
converges to a function of the form
$f((1-x)\tilde{k}^m)$ for large $m$ ($m<<d$), and $f(x)$ satisfies the 
Poincar\'{e} functional relation: 
\begin{equation}
G_{1}(f(y))=f(y \tilde{k}),
\end{equation}
where $y=1-x$. The function form of $f(y)$ can be uniquely 
determined from Eq. (2). 

The solution of $G_{1}(f_\infty)=f_\infty$ gives the probability that a link 
is not connected to the giant component of the network by one of its
ends \cite{braunstein}.
It is known \cite{bingham} that $f(x)$ has an asymptotic functional form, $f(y)=f_\infty+ay^{-\delta}+0(y^{\delta})$.
Expanding both sides of Eq. (2) we obtain:
\begin{equation}
G_{1}(f_{\infty})+G_{1}^{'}(f_{\infty})ay^{-\delta}=f_{\infty}+a\tilde{k}^{-\delta}y^{-\delta}+0(y^{\delta}).
\end{equation}
Since $G_{1}(f_{\infty})=f_{\infty}$, we have 
$\delta=-\ln G_{1}'(f_\infty)/\ln\tilde{k}$.

If $q(1)=0$ and $q(2)\neq 0$, from  $G_{1}(f_\infty)=f_\infty$, we 
have $f_{\infty}=0$ and 
$G^{'}_{1}(f_{\infty})=G_{0}^{''}(0)/\langle k\rangle=2q(2)/ \langle k \rangle$. 
If $q(2)=q(1)=0$ (B$\ddot{o}$ttcher case \cite{bingham}), then $\delta=\infty$,
which indicates that $f(y)$ has an exponential singularity. 
Therefore, networks with minimum degree $k_{m}\geq 3$ do not exhibit 
the following properties for $m<<d$, and therefore have no fractal boundaries.

Applying Tauberian like theorems \cite{bingham,wess} to $f(y)$, which has a 
power-law behavior for $y\to \infty$, Dubuc \cite{dubuc} concluded
that the Taylor expansion coefficient of $\tilde{G}_{m}(x)$, 
$\tilde{P}(B_{m})$, behaves
as $B_{m}^{\mu}$ with an exponential cutoff at $B_{m}^{*}\sim \tilde{k}^{m}$.
When $q(1)\neq 0$ and $q(2)\neq 0$, we have $\mu=\delta-1$ and when $q(1)=0$ and $q(2)\neq 0$, 
we have $\mu=2\delta-1$.
Thus the distribution of the number of nodes in the shell $m$ with $m<<d$ has a power law tail for 
small values of $B_{m}$: 
\begin{equation}
\tilde{P}(B_{m})\sim B_{m}^{\mu}.
\end{equation}

For ER network, Eq. (4) is supported by simulations for $m\leq d$ in Fig.~\ref{fig3}c.

The above considerations are correct only for $m<d$, for which the depletion of nodes with
large degree in the network is insignificant.

In a large network, the shells with $m>>1$ behave almost deterministically and 
there exists a functional relation between
any two shell $m$ and shell $n$ with $n>m$ (a detailed proof will be given elsewhere):
\begin{equation}
r_{n}=G_{0}(G_{1}^{n-m}(G_{0}^{-1}(r_{m}))),
\end{equation}
where $r_{n}$ is the fraction of nodes outside shell $n$. It can also be shown that the branching
factor in the $r_n$ fraction of nodes is $\tilde{k}(r_n)=uG_{0}^{''}(u)/G_{0}^{'}(u)$,
where $u=G_{0}^{-1}(r_n)$.

For ER networks, Eq. (5) yields:
\begin{equation}
r_{\ell+1}=e^{\langle k \rangle(r_{\ell}-1)}=\Sigma _{\ell=0}^{\infty} q(k)r_{\ell}^{k},
\end{equation}
which is valid for all possible $\ell$. We test it in Fig.~\ref{fig3}d.

When $m<<d$ and $n>>d$, using the same considerations as before it can be 
shown that:
\begin{equation}
r_n=[a\tilde{k}(1-r_m)]^{-\mu-1}+r_{\infty},
\end{equation}
where $r_{\infty}=G_{0}(f_{\infty})$ is the fraction of nodes not belonging to the giant component of the
network, $a$ is a constant.

Based on Eqs. (4) and (7),
expressing $r_{m}$ and $r_{n}$ in terms of $B_{m}$ and $B_{n}$, we find that for $m<<d$ and $n>>d$, 
$B_{n}\sim B_{m}^{-\mu-1}$.
 Using $\tilde{P}(B_n)dB_n=\tilde{P}(B_m)dB_m$, we obtain
\begin{equation}
\tilde{P}(B_n)\sim B_n^{-1-\mu /(\mu+1)-1/(\mu+1)}=B_{n}^{-2}, 
\end{equation}
supporting the 
numerical findings in Fig.~\ref{fig2}. 

These results are rigorous when $\tilde{k}$ exists and when the
minimum degree $k_{m} \leq 2$. For SF networks with $\lambda <3$,  
$\tilde{k}$ diverges for $N\to \infty$. But for finite $N$, $\tilde{k} $ still exists.
Thus the above results can also be applied to the case of $\lambda <3$. 
For both ER and SF networks with $k_{m}\geq 3$, the power law of 
$P(B_n)$ with $n>>d$ cannot be observed, as we indeed confirm by simulations.

Relating our problem with percolation theory, we can explain the simulation 
results of probability distribution of cluster size $s_\ell$.
The cluster size distribution in percolation at some concentration $p$ close to $p_{c}$
is determined by the formula \cite{cohen}:
\begin{equation}
P_p(s>S)\sim S^{-\tau+1}\exp(-S|p-p_c|^{1/\sigma})\;.
\end{equation}
In the case of random networks 
the percolation threshold is given by $p_{c}=1/ \tilde{k}$.
In the exterior of the shell $n$ ($n>>d$),
we can estimate $|p-p_{c}|\sim (\tilde{k}(r_{n})-1)/\tilde{k}$, 
where $\tilde{k}(r_{n})$ decreases and reaches the critical percolation value of $1$.

The cluster size distribution can be estimated 
by introducing a sharp exponential cutoff at 
$s=S_{n}^{*}\sim |\tilde{k}(r_{n})-1|^{-\frac{1}{\sigma}}$, so that
$P_{n}(s>S)\sim S^{-\tau+1}P(S_{n}^{*}>S)$, where $P(S_{n}^{*}>S)$ is the
probability for a given shell to have $S_{n}^{*}>S$.

Since $r_{n}-r_{\infty}$ has a smooth power law distribution and $\tilde{k}(r_{\infty})<1$,
the probability that $|\tilde{k}(r_{n})-1|<S^{-\sigma}= \varepsilon$ 
is proportional
to $\varepsilon$. Thus $P(S_{n}^{*}>S)\sim S^{-\sigma}$ and $P_{n}(s>S)=S^{-\tau+1-\sigma}$ \cite{barabasiPRL}.
Therefore the cluster size distribution follows $n(s)\sim s^{-(\tau +\sigma)}$. 

For ER networks and SF networks with $\lambda >4$,
$\tau=2.5$ and $\sigma=0.5$,
the above derivations lead to $n(s)\sim s^{-3}$. 
For SF networks  
with $2<\lambda <4$,
$\tau=(2\lambda-3)/(\lambda-2)$ and $\sigma=| \lambda-3 |/(\lambda-2)$ \cite{cohenpre}. 
Thus, for SF network with $\lambda >3$, there will be 
$n_{s}\sim s^{-3}$. We conjecture $n_{s}\sim s^{-3}$ 
even for $2<\lambda <3$, although
in this case $\tilde{k}(r_{n})$ does not exist and the above derivations are not valid. Our
numerical simulations support these results in Fig.~\ref{fig4}a, b.

In summary, we find empirically and analytically that the boundaries 
of a broad class of complex networks including non-fractal networks
\cite{song1} have fractal features.
Our findings can be applied to the study of epidemics.
It implies that a strong decay of the epidemic will happen in the boundaries of human network,
due to the low degree of nodes. 
The fractal clusters at the boundaries, which are connected sparsely to the bulk, 
may explain why big breakout of 
epidemical disease (such as the ``black death'' in medieval Europe) would
suddenly stop after affecting a large percentage of population.

We thank ONR and Israel Science
Foundation for financial support.

\vspace*{-0.3cm}


\begin{thebibliography}{99}
\bibitem{similar} Different realizations yield similar results.
              In one realization, a certain fraction of nodes are randomly taken to be origin. 
                   The histogram is obtained from $B_{\ell}$ belonging to different origin nodes.
\bibitem{fig3c} The behavior of the pdf of $B_{\ell}$ for $\ell<d$ will
                be discussed later and is shown in Fig.~\ref{fig3}c.
\bibitem{yeast} We also find similar results (not shown here) for other real networks.
\bibitem{fraction} Fractional shells with $\ell+a$ ($0<a< 1$) are used here by 
                removing nodes within shell $\ell$ and a fraction $a$ of nodes at shell $\ell$.
\end{thebibliography}
\end{document}